\documentclass[12pt,aps]{revtex4}
\usepackage[dvips]{graphicx}
\begin{document}

\begin{center}
{\large\bf $C^{12}$ shape isomers in the chiral field solitons approach. }\\[5mm]
{\large V.A. Nikolaev$^1$}\\
    {\small NTL, INRNE, Sofia , Bulgaria}\\[3mm]
{\large Yu.V. Chubov$^2$, O.G. Tkachev$^3$}\\
    {\small Institute of Physics and Information Technologies},\\
    {\small Far East State University, Vladivostok, Russia.}\\[10mm]
{\it E-mail:
    $^{1}$nikolai@spnet.net\ \ \
    $^{2}$yurch@ifit.phys.dvgu.ru\ \ \
    $^{3}$tkachev@ifit.phys.dvgu.ru}
\end{center}

\vspace{1cm}
\begin{center}
\parbox{14cm}{
\centerline{\bf Abstract} {\small The variational approach to the
problem of seeking axially symmetric solitons with $B=12$ is
presented. The numerically obtained local minima of the skyrmion
mass functional and baryon charge distributions are pointing to
the possible existence of shape isomers in $C^{12}$ spectra in the
the framework of the original Skyrme model. Theoretical analysis
reveals the exclusiveness of each individual state manifested in
the structure of the solitons  from the given topological sector
$B=12$.}}
\end{center}

\vspace{1cm}
\section{Introduction.}
     The  Skyrme model\cite{Skyrme} was proposed in the sixties
  as a model for the strong interactions of hadrons and was
  very successful in describing nucleons as quantum states of the chiral soliton in original and generalized Skyrme model \cite{NikNov}.
     The Skyrme model gives us very  unusual instrument to  study new physics
     especially in the light nuclei region. In this region  traditional one nucleon degrees of freedom are possibly
     not so important as  finite sizes of nucleon, comparable to the nuclear
     radiuses \cite{Nik}, \cite{hep-ph0109192}.
There is no  analytic solutions for the Skyrme model equations of
motion. We still have to use  variational approaches. The most
popular in between them is the so called rational map ansatz
\cite{ Mant} leading to the a number of the  solution with
discrete space symmetries and topological charges corresponding to
light nuclei atomic numbers up to 22. They  are very like to
fullerene structures more usual for the larger molecular scale
\cite {Battye}. In any way such solutions are like pure numerical
solutions obtained in \cite{Braaten0} for topological charges
2,3,4,5 and 6.

    Here we try to search soliton with axially symmetric baryon charge
    distribution. Quantization procedure for the states with baryon number equal to 2,3 and 4 was worked out
   in \cite{Weigel} -\cite{Kozhevnikov} without vibrations  have been taken  into account  and including the breathing
mode\cite{Nikolaev1} -\cite{Nikolaev2}. The \cite{Nik.Sov.23} and
\cite {NikJaf} describe quantization rules for axially symmetric
soliton we are considering here.

     The variational ansatz we use here was proposed independently in
\cite{Nikolaev3},\cite{Kurihara} and \cite{Sorace}). The ansatz
being very simple, gives the possibility to do  analytical
analysis of a part of the nuclear problem.

    In this paper we present the results of our variational calculations of
the classical soliton structure with baryon charge $B=12$ in the
framework of the original $SU(2)$ Skyrme model. After the
quantization procedure some of these solitons could  be identified
with shape isomers of $C^{12}$.

\section{Ansatz for the Static Solutions.}
 We follow our papers \cite{Nikolaev5} and \cite{Nik24} with some modifications. In
variational form
 of the chiral field $U$:
\begin{eqnarray}
U(\vec r ) = cosF(r) + i(\vec {\tau} \cdot \vec N )\ sinF(r) .
\label{e1}
\end{eqnarray}
we use the next  general assumption about the configuration of the
isotopic vector field $\vec N$ for axially symmetric soliton:
\begin{eqnarray}
\vec N = \{ cos(\Phi (\phi ,\theta ))\cdot sin(T(\theta )),
\ sin(\Phi (\phi ,\theta ))\cdot sin(T(\theta )),\ cos(T(\theta )) \} .
\label{e2}
\end{eqnarray}
In eq.(\ref{e2}) $\Phi (\phi),\ T(\theta)$ are some arbitrary functions of
angles $(\theta ,\phi)$ of the vector $\vec r$ in the spherical coordinate
system.

\section{Mass Functional and Solutions for Static Equations.}
After some algebra  (\ref{e1}), (\ref{e2}) and  the Lagrangian
density ${\cal L}$ for the stationary solution
\begin{eqnarray}
{{\cal L}} = {F_\pi^2\over 16}\cdot Tr(L_k L_k)+{1\over
32e^2}\cdot Tr\Bigl[ L_k,L_i\Bigr ]^2 ,\label{e3}
\end{eqnarray}
expressed through the left currants $L_k= U^+{\partial}_kU$  lead
to the expression
\begin{eqnarray}
{{\cal L}} = {{\cal L}}_2 + {{\cal L}}_4 ,\label{e4}
\end{eqnarray}
where
$${\cal L}_2= -{F_\pi^2\over 8}\left\{\left({\partial F\over\partial x}\right)
^2\hspace{-0.2cm}+\left[\left({sinT\over sin\theta}{\partial\Phi\over\partial
\phi}\right)^2\hspace{-0.2cm}+\left({\partial T\over\partial\theta}\right)^2
\hspace{-0.2cm}+sin^2T\left({\partial\Phi\over\partial\theta}\right)^2\right]
{sin^2F\over r^2}\right\}$$
and
\begin{eqnarray}
{\cal L}_4 &=& -{1\over{2e^2}}\cdot{sin^2F\over r^2}\cdot\Biggl\{{sin^2T\over
sin^2\theta}\left({\partial T\over\partial\theta}\right)^2\left({\partial\Phi
\over\partial\phi}\right)^2\cdot{sin^2F\over r^2}\nonumber\\
&+&\left[{sin^2T\over sin^2\theta}\left({\partial\Phi\over\partial\phi}\right)
^2+\left({\partial T\over\partial\theta}\right)^2+sin^2T\left({\partial\Phi
\over\partial\theta}\right)^2\right]\left({\partial F\over\partial x}\right)^2
\Biggr\}\label{e6}
\end{eqnarray}

     The variation of the functional $L=\int {\cal L}d\vec r$ with respect to
$\Phi$ leads to an equation which has a solution of the type
\begin{eqnarray}
\Phi(\phi) = k(\theta)\cdot\phi+Const\label{e7}
\end{eqnarray}
with a constrain:
\begin{eqnarray}
{\partial\over\partial\theta}\bigl[sin^2T(\theta)\cdot sin\theta\cdot
{\partial k(\theta)\over\partial\theta}\bigr]= 0.\label{e8}
\end{eqnarray}
It is easily seen from eq.\cite{Nikolaev5} that function
$k(\theta)$ may be piecewise constant function (step function) in
general case:
\begin{eqnarray}
\Phi(\theta ,\phi) = \left\{
  \matrix{
    \displaystyle{k^{(1)} \phi+\rho^{(1)}\ ,
          \ \ {\rm for}\ 0\leq\theta < \theta_1\ ,}\hfill\cr\cr
    \displaystyle{k^{(2)} \phi+\rho^{(2)}\ ,
          \ \ {\rm for}\ \theta_1\leq\theta < \theta_2\ ,}\hfill\cr\cr
    \displaystyle{\ .\ \ \ \ .\ \ \ \ .}\hfill \cr\cr
    \displaystyle{k^{(l)} \phi+\rho^{(l)}\ ,
           \ \ {\rm for}\ \theta_{l-1}\leq\theta < \pi\ .}\hfill
  }\right.\nonumber
\end{eqnarray}
Moreover $k(\theta)$ must be integer in any region
$\theta_m\leq\theta\leq\theta_{m+1}$, where $\theta_m$,
$\theta_{m+1}$ are successive points of discontinuity. The
positions of these points are determined by the condition
\begin{eqnarray}
T(\theta_m) = m\cdot\pi,\label{e9}
\end{eqnarray}
with integer $m$, as follows from eq.(\ref{e8}).

     Now we have the following expression for the mass of the soliton
\begin{eqnarray}
M = \gamma\cdot\bigr[a\cdot A+b\cdot B+C\bigr],\label{e10}
\end{eqnarray}
where $\gamma = \pi\cdot F_\pi /e$ and $x = F_\pi \cdot e\cdot r$ and the $a,b$
and $A,B,C$ are the following integrals:
\begin{eqnarray}
a =\int\limits_{0}^{\pi}\Bigl[k^2{sin^2T\over sin^2\theta}+(T^\prime )^2\Bigr]
sin\theta d\theta,\ \ b=k^2\int\limits_{0}^{\pi}{sin^2T\over sin^2\theta}
(T^\prime)^2 sin\theta d\theta,\label{e11}
\end{eqnarray}
\begin{eqnarray}
A=\int\limits_{0}^{\infty}sin^2F\Bigl[{1\over 4}+(F^\prime)^2\Bigr]dx,\
B=\int\limits_{0}^{\infty}{sin^4F\over x^2}dx,\ C={1\over 2}\int\limits_{0}
^{\infty}(F^\prime x)^2dx.\label{e12}
\end{eqnarray}
Here we use the symbol prime to denote the following derivatives
\begin{eqnarray}
\Phi^\prime = {\partial\Phi\over \partial\phi} \ \ ; \ \ T^\prime = {\partial T
\over\partial\theta} \ \ ;\ \ F^\prime ={\partial F\over \partial r}\label{e13}
\end{eqnarray}

     We consider the configurations with finite masses. The only configurations
which obey the finiteness of mass condition are the configurations with
$F(0)=n\cdot\pi$ where $n$-is some integer number. Without loss of generality
we take $F(\infty )=0$. As it was shown in\cite{Nikolaev5} $T(\theta )$ has the
following behaviour near the boundary of the domain of its definition
\begin{eqnarray}
T(\theta )\to\theta^k,\ for\ {\theta\to 0}\ ;\ \ T(\theta )\to\pi\cdot l-(\pi -
\theta )^k,\ for\ {\theta\to \pi} .\label{e14}
\end{eqnarray}
Here $l$ is an integer number. Thus we have the following estimation for the
number of discontinuity points $d$:
\begin{eqnarray}
0\leq d\leq l-1.\label{e15}
\end{eqnarray}
Now all solutions $U_{l\{k_i,n_i\}}$ are classified by a set of integer numbers
$l$, $k_0,...,k_{l-1}$ and $n_0,...,n_{l-1}$. The functions $F(x)$ and $T(\theta)$ have to obey
the equations (14,15) from\cite{Nikolaev5} in arbitrary space region with given
number $k$.

\section{Baryon Charge Distribution and the Soliton Structure.}
Now consider more carefully the structure of solitons. For that purpose
let us calculate the baryon charge density
\begin{eqnarray}
J_0^B(\vec r) = - {1\over 24\pi^2}\cdot\epsilon_{0\mu\nu\rho} Tr( L_\mu L_\nu
L_\rho ).\label{e18}
\end{eqnarray}
The straightforward calculation gives
\begin{eqnarray}
J_0^B(r,\theta )=-{1\over 2\pi^2}\cdot{sin^2F\over r^2}\cdot{dF\over dr}\cdot
{sinT\over sin\theta}\cdot{dT\over d\theta}\cdot{d\Phi\over d\phi}\label{e19}
\end{eqnarray}
Equation (\ref{e19}) immediately results in the expression for the
corresponding topological charge
\begin{eqnarray}
B = -\sum\limits_{m=0}^{l-1}(-1)^m\cdot  n_m\cdot k_m\label{e20}
\end{eqnarray}

In\cite{Nikolaev5} we have investigated toroidal multiskyrmion
configurations with baryon numbers $B=1,2,3,4,5$ and more
complicated nontoroidal (including antiskyrmions ($\bar S$))
configurations.

It is obvious that setting $k_m<0$ for even m, $k_m>0$ for odd m and $n_m>0$ for all m,
we obtain configuration with positive baryon charge. In the general case for obtaining
configuration with positive baryon charge we must require that
$$n_m\cdot k_m > 0\ \ \ for\ odd\ l\ ,$$
$$n_m\cdot k_m < 0\ \ \ for\ even\ l\ .$$

\section{The Masses and baryon charge distributions.}
Here we reproduce the mass values and baryon charge distributions
corresponding to the obtained local minima of the energy
functional for the Skyrme field. We have to point out that  we
discuss multiskyrmion configurations we search for  not only
classically stable configurations (The decay in two or more
skyrmions is forbidden energetically). Nonstable configurations
are also in our attention because they may become stable after the
quantization procedure \cite{Nikolaev5} or pion field Casimir
energy would taken into account in full quantum description of the
considered solitons.

We restrict ourself to configuration with a symmetric distribution
of energy (mass) density in the ($x,y$) plane. This mean that from
the class of all solutions considered, characterized by the
numbers $l, \{k_m, n_m\}\vert_1^l$, we choose only the solution
satisfying the condition
\begin{eqnarray}
&&k_i=k_{l+1-i},\ \ \ n_i=n_{l+1-i}\ \ \ {\rm for\ odd}\ l,\cr
&&k_i=-k_{l+1-i},\ \ \ n_i=-n_{l+1-i}\ \ \ {\rm for\ even}\ l.\label{eee}
\end{eqnarray}

In Table1 we present the masses of shape isomers of $C^{12}$. The
calculated soliton masses are given in $(\pi F_\pi/e)$ units.

\begin{table}[htb]
\begin{center}
\begin{tabular}{| c | c |}
\hline
Configuration & Mass $(\pi F_\pi/e)$\\
\hline
2\{3.2-3.2\}        & 157.6778  \\
2\{6.1-6.1\}        & 137.9763  \\
3\{2.2-2.2-2.2\}    & 172.3576  \\
3\{2.2-4.1-2.2\}    & 161.6704  \\
3\{4.1-2.2-4.1\}    & 145.1425  \\
3\{4.1-4.1-4.1\}    & 134.4552  \\
\hline
\end{tabular}
\end{center}
\caption{$C^{12}$ soliton mass spectrum.}
\end{table}

From Table 1 one see that in calculated part of spectrum  all
configurations have very different structures. Presence of such
isomers could probably be seen in high energy ion-ion scattering
experiments.

\begin{figure}[htp]
\centerline{ \includegraphics[width=155mm]{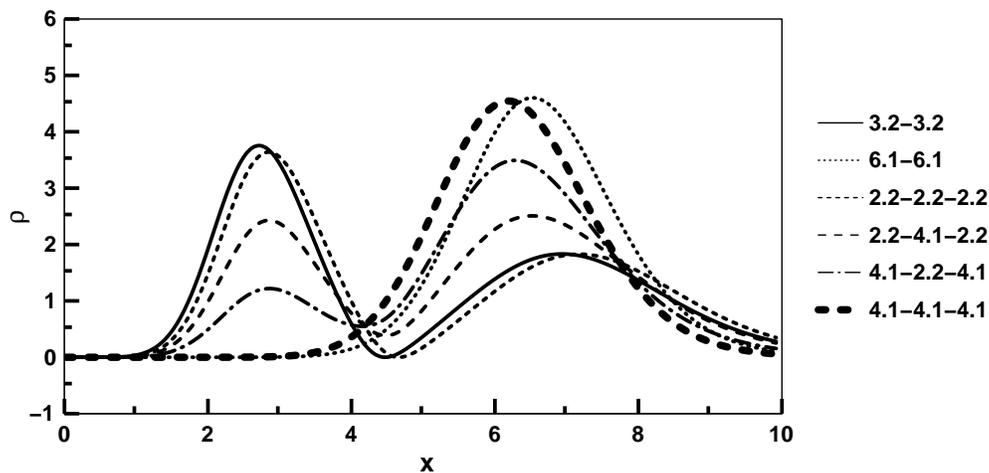}}
\vspace*{-10mm} \caption{Baryon density of $C^{12}$ shape
isomers.}
\end{figure}

The reason we are looking for possible not pure toroidal  solitons
is that there is a number of solutions with smaller masses among
them. For example a configuration composed from three toroidal
multibaryons 3\{4.1-4.1-4.1\} has a smaller mass. Now this state
can not decay into 12 classical skyrmions with $B=1$
($M_{1\{1.1\}}=11.60608\pi F_\pi/e$) or into three toroidal
skyrmion with $B=4$ ($M_{1\{4.1\}}=47.67478\pi F_\pi/e$). We point
such a skyrmion  as classically stable configuration. The
configuration 3\{2.2-2.2-2.2\} do not obey the condition of
classical stability.

For calculated  axially symmetric configurations we present baryon
density distributions integrated on $d\Omega=\sin(\theta)d\phi
d\theta$ at Figure 1. Here we use dimensionless coordinates
$x=F_\pi e r$.

In according to our calculations the solitons from the same
topological sector  can have strongly different masses if they
have different structure. We also have to point out that a number
of the states have shell like structure. There are states which
has $n_i\neq 1$. Such shape isomer can give different specific
contribution to physics processes in light nuclei.

\section{Conclusions.}

\vspace{.2cm} The axially symmetric solitons with baryon number $B
= 12$ have been investigated in the framework of the very general
assumption about the form of the solution of the Skyrme model
equations. The obtained solitons could be seen in nuclear
reactions as isomer contributions in  reactions involving
$C^{12}$. Such isomers correspond to  different form of baryon
density distribution. We have to point out that used ansatz leads
to stable solitons with $B=12$ and shell like structure of the
baryon density distribution.

\section{Acknowledgements.}
\noindent
We thank A.N. Antonov and V.K. Lukyanov for useful discussions of obtained results.\\
This work is supported in part by the programme "University of Russia" UR.02.01.020

\end{document}